%% file: ASICON2013.tex
\documentclass[9pt, conference, letterpaper]{IEEEtran}
 
\newcommand{\mbf}{\mathbf}
\newcommand{\wtd}{\widetilde}

\usepackage[ruled, linesnumbered]{algorithm2e}
\usepackage{url}
\usepackage{multirow}
\usepackage{graphicx}
\usepackage{amssymb}
\usepackage{amsmath}
\usepackage{amsfonts}
\usepackage{color}
\usepackage{epstopdf}
\usepackage{verbatim} 

\begin{document} 
\vspace{-0.05in}
\title{{Power Grid Simulation using Matrix Exponential Method \\with Rational Krylov Subspaces} }
\author{Hao~Zhuang,~\IEEEmembership{Student Member,~IEEE,}Shih-Hung~Weng,~\IEEEmembership{Student Member,~IEEE,}and~Chung-Kuan~Cheng\IEEEmembership{Fellow,~IEEE}\\
        Department of Computer Science and Engineering, University of California, San Diego, CA, 92093, USA\\
        hao.zhuang@cs.ucsd.edu, s2weng@ucsd.edu,~ckcheng@ucsd.edu
        }
\maketitle
\vspace{-3.0in}
\begin{abstract}
One well-adopted power grid simulation methodology is to factorize matrix once and perform 
only backward/forward substitution with a deliberately chosen step size along the simulation. 
Since the required simulation time is usually long for the power grid design, the costly 
factorization is amortized. However, such fixed step size cannot exploit larger step size for 
the low frequency response in the power grid to speedup the simulation. In this work, we utilize 
the matrix exponential method with the rational Krylov subspace approximation to enable adaptive 
step size in the power grid simulation. The kernel operation in our method only demands one factorization 
and backward/forward substitutions. Moreover, the rational Krylov subspace approximation can relax the 
stiffness constraint of the previous works~\cite{Weng11_ASICON}\cite{Weng12_TCAD}. 
The cheap computation of adaptivity in our method could exploit the long low-frequency response in a 
power grid and significantly accelerate the simulation. The experimental results show that our method 
achieves up to 18X speedup over the trapezoidal method with fixed step size. 
\end{abstract}
\vspace{-0.08in}
\section{Introduction}
\label{sec:intro}
\input{intro}

\vspace{-0.08in}
\section{Preliminary}
\label{sec:preliminary}
\vspace{-0.08in}
\subsection{Power Grid Formulation}
\label{sub-sec:pwr_formulation}
\input{sub-pwr_grid_formulation}
\vspace{-0.1in}
\subsection{Matrix Exponential Method}
\label{sub-sec:mexp_method}
\input{sub-mexp_method}

\vspace{-0.08in}
\section{MEXP with Rational Krylov Subpsace}
\vspace{-0.05in}
\label{sec:rational_krylov}
\input{sub-rational_krylov}

\vspace{-0.08in}
\section{Adaptive Time Step Control}
\vspace{-0.05in}
\label{sec:adaptive_step_control}
\input{adaptive_step_control}

\section{Experimental Results}
\vspace{-0.05in}
\label{sec:exp_results} 
\input{sub-exp_results}

\vspace{-0.1in}
\section{Conclusion}
\label{sec:conclusion}
\input{sub-concl}

\section{Acknowledgements}
This work was supported by NSF CCF-1017864.
\bibliographystyle{abbrv}
\bibliography{ASICON2013}  
\vspace{-0.05in}
\end{document}

%% file: intro.tex
Power grid simulation is a very essential and computational heavy tasks during VLSI design.  
Given current stimulus and the power grid structure, 
designers could verify and predict the worst-case voltage noise 
through the simulation before signing off their design.  
However, with the huge size of modern design, power grid simulation is a time-consuming process. Moreover, manifesting effects 
from the package and the board would require longer simulation time, e.g., up to few $\mu s$, which worsens the performance 
of the power grid simulation. Therefore, an efficient power grid simulation is always a demand from industry. 

Conventionally, the power grid simulation is based on the trapezoidal method where the major computation is to solve a linear 
system by either iterative approaches or direct methods. 
The iterative methods usually suffer from the convergence problem 
because of the ill-conditioned matrix from the power grid design. On the other hand, the direct 
methods, i.e., Cholesky or LU factorizations, are more general for solving a linear system. Despite the huge memory 
demanding and computational effort, with a carefully chosen step size, the power grid simulation could perform only one 
factorization at the beginning while the rest of operations are just backward/forward substitutions. Since a power grid design 
usually includes board and package models, a long simulation time is required to manifest the low-frequency response. Hence, 
the cost of expensive factorization can be amortized by many faster backward/forward substitutions. Such general factorization 
and fixed step size strategy\cite{Xiong12}\cite{Yang12}\cite{Yu12}\cite{Zhao02} is widely adopted in industry. 

The matrix exponential method (MEXP) for the circuit simulation has better accuracy and adaptivity because of the 
analytical formulation and the scaling invariant Krylov subspace approximation\cite{Weng11_ASICON}\cite{Weng12_TCAD}. 
Unlike the fixed step size strategy, MEXP could 
dynamically adjust the step size to exploit the low-frequency response of the power grid without expensive computation. 
However, 
the step size in MEXP is limited by the stiffness of circuit. This constraint would drag 
the overall performance of MEXP for the power grid simulation. 

In this work, we tailor MEXP using \emph{rational Krylov subspace} for the power grid simulation with adaptive time stepping. 
The rational Krylov subspace uses $(\mbf{I}-\gamma \mbf{A})^{-1}$ as the basis instead of $\mbf{A}$
used in the conventional Krylov 
subspaces, where $\mbf I$ is an identity matrix and $\gamma$ is a predefined parameter. The rational basis limits the spectrum of a circuit system, and emphasizes small magnitude eigenvalues, which are important to exponential function, so that the exponential of a matrix can be accurately approximated 
.
As a result, MEXP with rational Krylov subspace can enjoy benefits of the adaptivity and the accuracy of MEXP. 
Even though the rational Krylov subspace still needs 
to solve a linear system as the trapezoidal method does, MEXP can factorize the matrix only once and then constructs the rest of 
rational Krylov subspaces by backward/forward substitutions. Therefore, MEXP can utilize its capability of adaptivity to accelerate 
the simulation with the same kernel operations as the fixed step size strategy. Overall, our MEXP enables adaptive time stepping for the power grid simulation with only one LU factorization, and allows scaling large step size without compromising the accuracy. 
The experimental results demonstrate the effectiveness of MEXP with adaptive step size. The industrial power grid designs can be accelerated 17X on average compared to the trapezoidal method. 

The rest of paper is organized as follows. Section~\ref{sec:preliminary} presents the background of the power grid simulation and MEXP. 
Sections~\ref{sec:rational_krylov} and~\ref{sec:adaptive_step_control} show the theoretical foundation of the rational Krylov subspace 
and our adaptive step scheme for the power grid simulation, respectively.  Section~\ref{sec:exp_results} presents experimental results 
of several industrial power grid designs. Finally, Section~\ref{sec:conclusion} concludes 
this paper.

%% file: sub-pwr_grid_formulation.tex
A power grid can be formulated as a system of 
differential equations via modified nodal analysis as below:
\begin{eqnarray}
    \label{eqn:ode}
    \mbf{C}\dot{\mbf{x}}(t) = - \mbf{G}\mbf{x}(t) + \mbf{B}\mbf{u}(t),
\end{eqnarray}
where matrix $\mbf{C}$ describes the capacitance and inductance, matrix 
$\mbf{G}$ represents the resistance and the incidence between voltages 
and currents, and matrix $\mbf{B}$ indicates locations of the independent 
sources. Vector $\mbf{x}(t)$ describes the nodal voltages and branch currents 
at time $t$, and vector $\mbf{u}(t)$ represents the corresponding supply voltage 
and current sources associated to different blocks. 
In this work, we assume those input sources are given and in the format of \emph{piece-wise linear}. 

%% file: sub-mexp_method.tex
MEXP\cite{Weng11_ASICON}\cite{Weng12_TCAD} is based on the analytical solution of~(\ref{eqn:ode}).
With initial solution 
from the DC analysis via direct\cite{Davis06} or iterative approaches\cite{Saad03}, the equation 
of MEXP from $t$ to $t+h$ can be expressed as
\begin{eqnarray} 
    \label{eqn:discrete_sol}
    \mbf{x}(t+h) = e^{\mbf{A}h}\mbf{x}(t) + \int^{h}_{0}e^{\mbf{A}(h-\tau)}\mbf{b}(t+\tau)d\tau.
\end{eqnarray}
where $\mbf{A} = - \mbf{C}^{-1}\mbf{G}$, and $\mbf{b}(t) = \mbf{C}^{-1}\mbf B\mbf{u}(t)$, when $\mbf{C}$ is not singular.
Assuming that input $\mbf{u}(t)$ is piece-wise linear (PWL), we integrate the last term in 
(\ref{eqn:discrete_sol}) analytically, turning the solution into the sum of three terms associated 
with matrix exponential operators,
\begin{eqnarray} 
\label{eqn:pwl_exact_sol}
    \mbf{x}(t+h)&=& e^{\mbf{A}h}\mbf{x}(t) \nonumber \\
		&+& (e^{\mbf{A}h} - \mbf{I})\mbf{A}^{-1}\mbf{b}(t) \nonumber \\
	        &+& (e^{\mbf{A}h}-(\mbf{A}h+\mbf{I}))\mbf{A}^{-2}\frac{\mbf{b}(t+h)-\mbf{b}(t)}{h}.
\end{eqnarray}
Eqn. (\ref{eqn:pwl_exact_sol}) has three matrix exponential terms, which are generally referred 
as $\varphi$-functions of the zero, first and second order~\cite{Niesen11}. It has been shown in 
\cite[Theorem 2.1]{Almohy11} that one can obtain the sum of them in one shot by computing the exponential 
of a slightly larger $(n+p)\times (n+p)$ matrix, where $n$ is the dimension of $\mbf{A}$, and $p$ is the 
order of the $\varphi$-functions ($p = 2$ in (\ref{eqn:pwl_exact_sol})). Thus, (\ref{eqn:pwl_exact_sol}) 
can be rewritten into
\begin{equation}
\label{eq:one_exp}
\mbf{x}(t + h) = \left[ 
			\begin{array}{*{20}{c}}
			    \mbf{I}_n & \mbf 0
			\end{array} 
		 \right]
		 e^{\mbf {\widetilde A} h}
		 \left[ 
			\begin{array}{*{20}{c}}
			    \mbf{x}(t)	\\
	    			\mbf    e_2
			\end{array} 
		\right],
\end{equation}
with
\begin{equation}
    \begin{gathered}
	\mbf {\widetilde A} = \left[ 
			 \begin{array}{*{20}{c}}
			    \mbf{A} & \mbf{W} \\
			    \mbf{0} & \mbf{J}
			\end{array} 
		   \right],\;\;\;
	\mbf{W} = \left[ 
			\begin{array}{*{20}{c}}
			    \frac{\mbf{b}\left(t + h \right) - \mbf{b}\left( t \right)}{h} & \mbf{b}\left( t \right)
			\end{array} 
		  \right] \hfill \\
	\mbf{J} = \left[ 
			\begin{array}{*{20}{c}}
			    0 & 1 \\
      			    0 & 0
			\end{array} 
		  \right],\;\;\;\;\;\;\;\;
\mbf e_2 = \left[ 
		    \begin{array}{*{20}{c}}
			0 \\
      			1
		    \end{array} 
	      \right] \hfill \\
    \end{gathered}
\end{equation}
To keep the notations simple, we use 
$\mbf{v}$ to represent 
$[\mbf{x}(t)\;\; \mbf e_2]^\mathsf{T}$ in the rest of paper, respectively. 
Note that the kernel computation of MEXP is to derive the exponential of a matrix, i.e., $e^\mbf{A}\mbf{v}$, 
which is approximated by the Krylov subspace method in works~\cite{Weng12_TCAD}\cite{Weng11_ASICON}. 
The Krylov subspace method has better scalability mainly from its sparse matrix-vector product centric computation. 
However, such approximation is only better for those eigenvalues with small magnitude, which 
means the maximum step size of MEXP in a stiff circuit has to be constrained in order to maintain the accuracy of 
approximation. In the following section, we will present how the rational basis could relax the stiffness constraint.

%% file: sub-rational_krylov.tex

In~\cite{Weng12_TCAD}\cite{Weng11_ASICON}, Eqn.~(\ref{eq:one_exp}) is calculated via the Krylov subspace method using Arnoldi process. The
subspace is defined as 
\begin{eqnarray}
    \label{eq:kry_A}
    \mbf {K_m}(\mbf{ \widetilde A, v}) = \text{span}\{\mbf v, \mbf { \wtd Av}, \cdots, \mbf {\wtd A}^{m-1} \mbf v \},
\end{eqnarray}
where $\mbf{v}$ is an initial vector. The Arnoldi process approximates the eigenvalues with large magnitude well. But when handling a stiff 
circuit system, the formed matrix usually contains many eigenvalues with small magnitude. Besides, $e^{\mbf{\wtd A}h}$ is mostly determined 
by the eigenvalues with smallest magnitudes and their corresponding invariant subspaces. In this scenario, due to the existence of 
eigenvalues with large magnitude in $\mbf {\wtd A}$, the Arnoldi process for Eqn.~(\ref{eq:kry_A}) requires large $m$ to capture the important 
eigenvalues (small magnitudes) and invariant spaces for exponential operator. Therefore, the time steps in MEXP has to be small enough to capture 
the important eigenvalues. This suggests us transforming the spectrum to intensify those eigenvalues with small magnitudes and corresponding invariant 
subspaces. We make such transformation based on the idea of \emph{rational Krylov subspace method}\cite{Moret04}\cite{Van06}. The details are
presented in the following subsections.
\vspace{-0.08in}
\subsection{Rational Krylov Subspaces Approximation of MEXP}
For the purpose of finding the eigenvalues with smallest magnitude first, we uses a preconditioner $ (\mbf I - \gamma \mbf{\wtd A})^{-1} $,
instead of using $\mbf {\wtd A}$ directly. It is known as the rational Krylov subspace\cite{Moret04}\cite{Van06}. The formula for the 
rational Krylov subspace is 
\begin{eqnarray}
    \label{eq:rat_kry}
    \mbf {K_m}((\mbf I- \gamma \mbf {\wtd A} )^{-1}, \mbf v)   =   
	    \text{span}\{ \mbf v, (\mbf I- \gamma\mbf {\wtd A} )^{-1}\mbf v,
    \cdots,  \nonumber \\
      (  \mbf I -  \gamma \mbf{\wtd A})^{-(m-1)} \mbf v \},
\end{eqnarray}
where $\gamma$ is a predefined parameter.  The Arnoldi process 
constructs $\mbf{V_{m+1}}$ and $\mbf H_{m+1,m}$, and the 
relationship is given by
\begin{eqnarray}
    \label{eq:rat_kry2}
    (\mbf{I}-  \gamma \mbf {\wtd A})^{-1} \mbf{V_m} =
    \mbf{V_m}\mbf H_{m,m} + h_{m+1,m} \mbf v_{m+1} \mbf e^{\mathsf{T}}_m, 
\end{eqnarray}
where $\mbf e_m$ is the $m$-th unit vector. Matrix $\mbf H_{m,m}$ is the
first $m \times m$ square matrix of an 
upper Hessenberg matrix of $\mbf H_{m+1,m}$, 
and $h_{m+1,m} $ is its last entry. 
$\mbf{V_m}$ consists of $[\mbf v_1, \mbf v_2, \cdots , \mbf v_m]$ 
, and 
$v_{m+1}$ is its last vector.
After re-arranging (\ref{eq:rat_kry2}) and given a time step $h$, 
the matrix exponential $e^{\mbf {\wtd A} h} \mbf v$ can be calculated as   
\begin{eqnarray}
    \label{eq:scale}
    e^{\mbf{\wtd A}h} \mbf{v} & \approx & \mbf{V_m}\mbf{V_m}^\mathsf{T} e^{\mbf{\wtd A}h}\mbf v 
			     = \left\|v\right\|_2\mbf{V_m}\mbf{V_m}^\mathsf{T} e^{\mbf{\wtd A}h}\mbf{V_m} \mbf e_1  \nonumber \\
			      & = & \left\| v \right\|_2 \mbf{V_m} e^{\alpha\mbf{{\widetilde H}_{m,m}}} \mbf e_1,
\end{eqnarray}
where $\mbf {\widetilde{H}}_{m,m} = \mbf I - \mbf H_{m,m}^{-1}$, $\alpha =\frac{h}{\gamma}$ is 
the adjustable parameters for control 
of adaptive time step size in Section \ref{sec:adaptive_step_control}.  Note that in practice, instead of computing 
$(\mbf I - \gamma \mbf{A})^{-1}$ directly, we only need to solve $(\mbf{C}+\gamma\mbf{G})^{-1}\mbf{C}\mbf{v}$, which can be achieved by one 
LU factorization at beginning. Then the construction of the following subspaces is by backward/forward substitutions.

This strategy is also presented in~\cite{Moret04}\cite{Van06}. 
Intuitively, the ``shift-and-invert'' operation would intensify the eigenvalues with small magnitudes and minify the 
eigenvalues with large magnitudes. By doing so, the Arnoldi process could capture those eigenvalues important to 
the exponential operator, which originally cannot be manifested with small $m$ in the conventional Krylov subspace. 
We would like to point out that the error bound for Eqn. (\ref{eq:scale}) does not longer depend on 
$\| \mbf {\wtd A}h \|$ as~\cite{Weng12_TCAD}.  It is only the first (smallest magnitude) 
eigenvalue of $\mbf {\wtd A}$. We observe 
that large $\alpha$ provides less absolute error under the same dimension $m$. An intuitive explanation is also given by \cite{Van06}, 
the larger $\alpha$ combined with exponential operators, the relatively smaller portion of the eigenvalues with smallest magnitude
determine the final vector. Within the assumption of piecewise linear in Eqn.~(\ref{eqn:pwl_exact_sol}),
our method can step forward as much as possible to accelerate 
simulation, and still maintain the high accuracy. The sacrifice resides in the small time step when more eigenvalues determine the final 
vector. So we should choose a appropriate parameter $\gamma$ or increase the order $m$ to balance the accuracy and efficiency. Even 
though the increasing $m$ results more backward/forward substitutions, the $m$ is still quite small in the power grid simulation. 
Therefore, it does not degrade our method too much. 

The formula of posterior error estimation is required for controlling adaptive step size.
We use the formula derived from \cite{Van06},
\begin{eqnarray}
    \label{eq:post_err_est}
    err(m,\alpha) 
    =  \frac{\left\| \mbf v \right\|_2}{\gamma} { h}_{m+1,m }
				 \left | 
				 (\mbf I - \gamma \mbf {\wtd{A}}) \mbf v_{m+1} 
				 \mbf e^\mathsf{T}_m 
				 \mbf H^{-1}_{m,m}
				 e^{  \alpha \mbf {  \widetilde{H}}_{m,m}} \mbf e_1 
				 \right | 
\end{eqnarray}
The formula provides a good approximation for the error trend with respect to $m$ 
and $\alpha$ in our numerical experiment.

\vspace{-0.08in}
\subsection{Block LU factorization}
In practical numerical implementation, in order to avoid 
direct inversion of $\mbf C$ to form $\mbf A$ in Eqn. (\ref{eq:rat_kry}),
the equation 
$ (\mathbf C  + \gamma \mathbf G )^{-1}\mathbf C
$
is used.
Correspondingly, for Eqn. (\ref{eq:one_exp}),
we uses the equations \begin{equation}
\label{eq:precond1}
    (\mathbf {\wtd C}  - \gamma \mathbf {\wtd G} )^{-1}\mathbf {\wtd C}
\end{equation}
where
$\mbf  {\wtd C} = 
    \left[
	\begin{array}{*{20}{c}} 
	    \mbf C & \mbf 0 \\ 
	 \mbf   0 & \mbf I
	\end{array}
    \right],
\mbf  {\wtd G} =
    \left[
	\begin{array}{*{20}{c}}
	 \mbf {-G}  & \mbf{ \wtd{W}} \\
	 \mbf 0 &  \mbf J
	\end{array}
    \right]
$, and
$\mbf{ \wtd{W} }
= \left[ 
\begin{array}{*{20}{c}}
\frac{\mbf{Bu}\left(t + h \right) - \mbf{Bu}\left( t \right)}{h} & \mbf{Bu}\left( t \right)
    \end{array} 
\right] $

The Arnoldi process based on Eqn.~(\ref{eq:precond1}) actually only requires to solve $\mbf v_{k+1}$ with $\mbf v_k$.
The linear system is expressed as
\begin{eqnarray}
   (\mbf {\wtd C } -  \gamma \mbf {\wtd G}) \mbf v_{k+1}=\mbf {\wtd C} \mbf v_k,
\end{eqnarray}
where $\mbf v_{k}$ and $\mbf v_{k+1}$ are $k$-th and $(k+1)$-th basis in the rational Krylov subspace. 
If $\mbf{\wtd W}$ changes with inputs during 
the simulation, the Arnoldi process has to factorize a matrix every time step. However, it is obvious that the majority of 
$\mbf {\wtd G}$
stay the same for this linear system. To take advantage of this property, a block LU factorization is devised here
to avoid redundant calculation.
The goal is to obtain the lower triangular $\mbf L$ and the upper triangular $\mbf U$ matrices: 
\begin{eqnarray}
    \label{eq:blu}
    \mbf  {\wtd C} - \gamma \mbf {\wtd G} = \mbf L \mbf U.
\end{eqnarray}
At the beginning of simulation, after LU factorization of 
$\mbf C + \gamma \mbf G=\mbf L_{sub} \mbf U_{sub} $,
we obtain the lower triangular sub-matrix $\mbf L_{sub}$, and upper triangular 
sub-matrix $\mbf U_{sub}$. Then Eqn. (\ref{eq:blu}) only needs updating 
via  
\begin{eqnarray}
\label{eq:blu1}
\mbf L = 
  \left[
    \begin{array}{*{20}{c}} 
       \mbf L_{sub} & \mbf 0\\
       \mbf 0 & \mbf I
    \end{array}
   \right] , \;\;
\mbf U = 
    \left[
    \begin{array}{*{20}{c}} 
	\mbf U_{sub} & -\gamma \mbf L^{-1}_{sub} \mbf {\wtd W}\\
	 \mbf 0 & \mbf I_J
    \end{array}
    \right],
\end{eqnarray}
where
$\mbf I_J =  {\mbf I - \gamma  \mbf J}$ is an upper triangular matrix. 
Assume we have $\mbf v_k$, 
the following equations further reduce 
operation $\mbf L_{sub}^{-1}$ and construct vector $\mbf{v}_{k+1}$:
$
\mbf z_1=[\mbf C, \; \mbf 0] \mbf v_k, \; 
\mbf z_2 = [\mbf 0, \; \mbf I] \mbf v_k; \;
\mbf y_2 = \mbf I^{-1}_J \mbf z_2, \; 
\mbf L_{sub} \mbf U_{sub} \; \mbf y_1 = \mbf z_1 + \gamma \mbf {\wtd W} \; \mbf y_2 \;  
$.
Then, we obtain 
$
\mbf v_{k+1} = [\mbf y_1, \; \mbf y_2]^{\mathsf{T}}$.
By doing this, it only needs one LU factorization at the beginning of simulation, and with 
cheap updates for the $\mbf L$ and $\mbf U$ at each time step during transient simulation. 


%% file: adaptive_step_control.tex
The proposed MEXP can significantly benefit from the adaptive time stepping because the rational Krylov subspace approximation 
relaxes the stiffness constraint as well as preserves the scaling invariant property. As a result, MEXP can effortlessly adjust 
the step size to different scale,
during the simulation. Such adaptivity is particularly 
helpful in the power grid where the voltage noise includes the high- to low-frequency responses from die, package and board. 

Our adaptive step scheme is to step forward as much as possible so that MEXP can quickly finish the simulation.  With the 
insight from Eqn. (\ref{eq:scale}), MEXP can adjust $\alpha$ to calculate results of required step sizes with only one 
Arnoldi process. However, even though the rational Krylov subspace could scale 
robustly, the step size in MEXP is restrained 
from input sources. As shown in Eqn.~(\ref{eqn:pwl_exact_sol}), MEXP has to guarantee constant slope during a stepping, and hence 
the maximum allowed step size $h$ at every time instant is limited. Our scheme will first determine $h$ from inputs at 
time $t$ and construct the rational Krylov subspace from $\mbf{x}(t)$. Then, $\mbf{x}$ within interval $[t,~t+h]$ are 
calculated through the step size scaling. 

Algorithm~\ref{algo:adaptive_step} shows MEXP with adaptive step control. In order to comply with the required accuracy during the 
simulation, the allowed error $err(m,\alpha)$
at certain time instant $t$ is defined
as $ err \le \frac{E_{Tol}}{T}h $
where $E_{Tol}$ is the error tolerance in the whole simulation process, $T$ is the simulation time span, $h$ is the 
step size at time $t$, and $err$ is the posterior error of MEXP from Eqn. (\ref{eq:post_err_est}). Hence, when 
we construct the rational Krylov subspace, we will increase $m$ until the $err(m,\alpha)$ satisfies the error tolerance. 
\begin{algorithm} 
    \label{algo:adaptive_step}
    \caption{MEXP with Adaptive Step Control}
    \KwIn{$\mbf{C}$, $\mbf{G}$, $\mbf{B}$, $\mbf{u}(t)$, $\tau$, error tolerance $E_{Tol}$ and simulation time $T$}
    \KwOut{$\mbf{x}(t)$}
    {
        $t = 0$; 
	$\mbf{x}(0)$ = DC\_analysis\;
        $[\mbf{L}_{sub},\mbf{U}_{sub}] = $LU($\mbf{C} +\gamma \mbf{G}$)\;
        \While{$t \leq T$}
        {	
	    Compute maximum allowed step size $h$ from $\mbf{u}(t)$ and
	    $\alpha = \frac{h}{\gamma}$ \;
	    Construct $\mbf{H_{m,m}}$, $\mbf{V_{m,m}}$, $err$ by Arnoldi process and (\ref{eq:post_err_est}) 
	    until $err(m,\alpha) \le \frac{E_{Tol}}{T}h$\;
	    Compute $\mbf{x}(t+h)$ by (\ref{eq:scale})\;
	    $t = t+h$; 
	}
    }
\end{algorithm}
The complexity of MEXP with adaptive time stepping is mainly determined by the total number of required backward/forward
substitutions
during the simulation process. 
The number of total substitution 
operations is 
$    \sum_{i=0}^N m_{i} $
where $N$ is total time steps, and $m_i$ is required dimension of the rational Krylov subspace at time step $i$. Compared 
to the trapezoidal method where the number of substitution operations depends only on the fixed step size, MEXP could use less 
substitution operations as long as the maximum allowed
step size $h$ is much larger than the fixed step size. Our 
experiments in the following section demonstrates it is usually the case for the power grid simulation.

%% file: sub-exp_results.tex
In this section, we compare performance of the power grid simulation 
by MEXP and the trapezoidal method (TR).
MEXP with adaptive step size control follows Algorithm~\ref{algo:adaptive_step}. 
We predefine $\gamma$ e.g. $10^{-10}$ here. 
and restrict the maximum allowed step size within $1ns$ to have enough time instants to plot the figure.
It is possible to have more fine-grain time instants, e.g., $10ps$, with only negligible cost by adjusting 
$\alpha$ in Eqn. (\ref{eq:scale}). TR is in fixed step size $h$ in order to minimize the cost 
of LU factorization. Both methods only perform factorization once for transient simulation, 
and rest of operations is mainly backward/forward 
substitution. We implement both methods in MATLAB and use UMFPACK package for LU factorization. Note that even 
though previous works\cite{Chen01_Efficient}\cite{Su03} show that using iterative 
approach in TR 
could also achieve adaptive step control, 
long simulation time span in power grid designs make direct 
method with fixed step size more desirable\cite{Xiong12}\cite{Yang12}\cite{Zhao02}. 
The experiments are performed on a Linux workstation with an Intel Core i7-920 2.67GHz CPU and 12GB memory.
The power grid consists of four metal 
layers: M1, M3, M6 and RDL. The physical parameters of each metal layer is listed in Table~\ref{tab:spec_metal_layers}. 
The package is modeled as an RL series at each C4 bump, and the board is modeled as a lumped RLC network. The
specification of each PDN design is listed in Table~\ref{tab:spec_pdn_designs} where the size of each design 
ranges from $45.7$K to $7.40$M. 
\vspace{-0.1in}
\begin{table}[htp]
  \caption{Widths and pitches of metal layers in the PDN design($\mu m$).}
  \centering
  \begin{tabular}{cccccccc}
    \multicolumn{2}{c}{M1} &\multicolumn{2}{c}{M3} &\multicolumn{2}{c}{M6} &\multicolumn{2}{c}{RDL}\\
    \hline pitch &width  &pitch &width  &pitch &width  &pitch &width \\
    \hline 2.5   &0.2  &8.5  &0.25  &30 &4 &400 &30\\
    \hline 
  \end{tabular}
  \label{tab:spec_metal_layers}
\end{table}
\vspace{-0.2in}
\begin{table}[htp]
    \caption{Specifications of PDN Designs}
    \centering
    \begin{tabular}{|c||c|r|r|r|r|}
    \hline
	Design  & Area ($mm^2$)	& \multicolumn{1}{|c|}{\#R}	    
				& \multicolumn{1}{|c|}{\#C}	
				& \multicolumn{1}{|c|}{\#L}	    
				& \multicolumn{1}{|c|}{\#Nodes}  \\ \hline
	D1	& $0.35^2$	& 23221	    & 15193	& 15193	    & 45.7K	\\ \hline
	D2	& $1.40^2$  	& 348582    & 228952	& 228952    & 688K	\\ \hline
	D3	& $2.80^2$  	& 1468863   & 965540	& 965540    & 2.90M	\\ \hline
	D4	& $5.00^2$  	& 3748974   & 2467400	& 2464819   & 7.40M	\\ 
    \hline
    \end{tabular}
    \label{tab:spec_pdn_designs}
\end{table}
\vspace{-0.1in}

In order to characterize a PDN design, designers can rely on the simulation result of impulse response of the 
PDN design. Many previous works\cite{Du10}\cite{Tian96} have proposed different PDN analysis based on the impulse 
response. The nature of impulse response of the PDN design, which contains low-, mid- and high-frequency components, 
can significantly enjoy the adaptive step size in MEXP.  We would also like to mention that the impulse response 
based analysis is not only for the PDN design, but also for worst-case eye opening analysis in the high speed interconnect
\cite{Shi08}. 

The impulse response can be derived from the simulation result of a step input from $0$V to $1$V with a small transition
time. Hence, we inject a step input to each PDN design and compare the performance of MEXP and TR. The transition time of 
the step input and the simulation time span is $10ps$ and $1\mu s$ for observing both high- and low-frequency responses. 
Table~\ref{tab:results_pdn_cases} shows the simulation runtime of MEXP and TR where the fixed step size is set as $10ps$ 
to comply with the transition time. In the table, ``DC'', ``LU'' and ``Time'' indicate the runtime for DC analysis, LU 
factorization and the overall simulation, respectively. DC analysis is also via the LU factorization. We can also adopt 
other techniques\cite{Xiong12}\cite{Yang12}\cite{Yu12} to improve the performance of DC analysis for both methods. 
It is noted that these cases are very stiff and 
with singular matrix $\mbf C$. 
We do not use the method\cite{Weng11_ASICON}\cite{Weng12_TCAD} on the benchmarks, 
because that it cannot
handle the singular $\mbf C$ in 
these industrial PDN design
without regularization.
It is worth pointing out that
even after regularization\cite{Weng12_TCAD}, the stiffness still causes large  $m$ series for matrix
exponential evaluation.
For example, we construct a simple RC mesh network with 2500 nodes. 
The extreme values of this circuit are 
$\mbf C_{min}=5.04 \times 10^{-19}$,
$\mbf C_{max}=1.00 \times 10^{-15}$,
$\mbf G_{min}=1.09 \times 10^{-2}$, 
and $\mbf G_{max}=1.00 \times 10^2$. 
The corresponding maximum eigenvalue 
of $-\mbf C^{-1}\mbf G$ is $-1.88 \times 10^{9}$
and minimum eigenvalue is $-3.98 \times 10^{17}$. 
The stiffness is 
$ \frac{Re(\lambda_{min})}{Re(\lambda_{max})}= 2.12 \times 10^{8}$ .
During simulation of $1ns$ time span, with a fixed step size $10ps$, MEXP based on~\cite{Weng11_ASICON}\cite{Weng12_TCAD} costs 
average and peak dimensions of Krylov subspace $m_{avg}$ = 115, and
$m_{peak}$=264, respectively. 
Our MEXP uses rational Krylov subspaces, which only need $m_{avg}$=3.11, $m_{peak}$=10
and lead to 224X speedup in total runtime.

\begin{figure}[ht]
    \centering
    \includegraphics[width=3.3in]{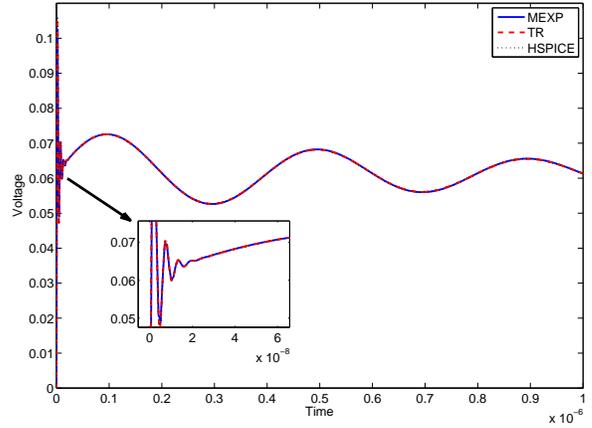}
    \caption{Result of D1}
    \label{fig:result_pdn_case1}
\end{figure}

In these test cases, our MEXP has significant speedup over TR because it can
adaptively exploit much large step size to simulate the design whereas TR can 
only march with $10ps$ time step for whole $1\mu s$ time span. The average speedup is 17X.  Fig.~\ref{fig:result_pdn_case1} shows the 
simulation result of design D1 at a node on M1. As we can see, the result by MEXP and TR are very close to the result of 
HSPICE, which is as our reference result here. The errors of MEXP and TR to HSPICE are $7.33\times10^{-4}$ and $7.47\times10^{-4}$.  
This figure also demonstrates that a PDN design has low-, mid- and high-freqeuncy 
response so that long simulation time span is necessary, meanwhile, small time steps are required during the 20$ns$ in the beginning. 
\vspace{-0.1in}
\begin{table}[htp]
    \centering
    \caption{Simulation runtime of PDN designs}
    \begin{tabular}{|c||r||r|r||r|r|r|}
    \hline
	\multirow{2}{*}{Design} & \multirow{2}{*}{DC(s)} &\multicolumn{2}{c||}{TR ($h=10ps$)} & \multicolumn{3}{c|}{Our MEXP ($\gamma=10^{-10}$) }  \\ 
	\cline{3-7} 
				 &	  &\multicolumn{1}{|c|}{LU(s)}   
					  &\multicolumn{1}{|c||}{Total}	
					  &\multicolumn{1}{|c|}{LU(s)}	 
					  &\multicolumn{1}{|c|}{Total}		
					  &\multicolumn{1}{|c|}{Speedup}  \\ \hline
        D1			 &0.710	  &0.670	   &44.9m	&0.680	 &2.86m		& 15.7 \\ \hline
        D2 			 &12.2	  &15.6  &15.4h	&15.5	 &54.6m	& 16.9 \\ \hline
        D3 			 &69.6   &91.6   &76.9h	&93.3	 &4.30h		& 17.9 \\ \hline
        D4 			 &219  &294  &204h	&299	 &11.3h	& 18.1	\\ 
    \hline
    \end{tabular}
    \label{tab:results_pdn_cases}
\end{table}

%% file: sub-concl.tex
For large scale power grid simulation, we propose an MEXP framework using two methods 
rational Krylov subspace approximation and adaptive time stepping technique. 
The former method can relax stiffness constraint of~\cite{Weng11_ASICON}\cite{Weng12_TCAD}. 
The later one helps adaptively exploit low-, mid-, and high-frequency property in   
simulation of industrial PDN designs.
In the time-consuming impulse response simulation, the proposed method 
achieve more than 15 times speedup on average over the widely-adopted fixed-step trapezoidal method.